\documentclass[
reprint,
amsmath,amssymb,
superscriptaddress,
onecolumn,
aps,
prfluids,
longbibliography,
floatfix
]{revtex4-2}

\usepackage[T1]{fontenc}
\usepackage[utf8]{inputenc}
\usepackage{graphicx}
\usepackage{bm}
\usepackage{booktabs}
\usepackage{dcolumn}
\usepackage[hidelinks]{hyperref}

\graphicspath{{figures/}}

\newcommand{\Oh}{\mathrm{Oh}}
\newcommand{\Bo}{\mathrm{Bo}}
\newcommand{\We}{\mathrm{We}}

\newcommand{\dd}{\mathrm{d}}
\newcommand{\vgamma}{v_\gamma}
\newcommand{\taugamma}{\tau_\gamma}
\begin{document}

\title{When do coalescing drops jump?}

\author{Vatsal Sanjay}
\email[]{vatsal.sanjay@comphy-lab.org}
\homepage[]{https://comphy-lab.org}
\affiliation{CoMPhy Lab, Department of Physics, Durham University, Science Laboratories, South Road, Durham DH1 3LE, United Kingdom}
\affiliation{Physics of Fluids Department, Max Planck Centre for Complex Fluid Dynamics, J. M. Burgers Centre for Fluid Dynamics, University of Twente, P.O. Box 217, 7500 AE Enschede, The Netherlands}

\author{Wilko Rohlfs}
\affiliation{Institute of Heat and Mass Transfer, RWTH Aachen University, Augustinerbach 6, 52056 Aachen, Germany}

\author{Detlef Lohse}
\email[]{d.lohse@utwente.nl}
\affiliation{Physics of Fluids Department, Max Planck Centre for Complex Fluid Dynamics, J. M. Burgers Centre for Fluid Dynamics, University of Twente, P.O. Box 217, 7500 AE Enschede, The Netherlands}
\affiliation{Max Planck Institute for Dynamics and Self-Organisation, Am Fassberg 17, 37077 G\"ottingen, Germany}

\date{\today}

\begin{abstract}
On a non-wetting substrate, the coalescence of two equal-sized liquid drops releases surface energy that may cause the coalesced drop to jump without external forcing, provided the drops are neither too viscous nor too heavy. Upon coalescence, a bridge forms between the two drops; its large neck curvature creates a Laplace-pressure gradient that draws liquid inward. For freely suspended drops, this motion would be redistributed symmetrically into shape oscillations. However, the substrate breaks that symmetry by blocking downward motion and redirects part of the momentum upward. Using three-dimensional direct numerical simulations and scaling arguments, we determine when this redirected motion results in take-off.
We find that the jumping/no-jumping transition is set by a capillary--viscous--gravity balance, namely by the late-recoil balance that also controls rebound inhibition of impacting drops. On the one hand, this balance explains the droplet arrest at a large enough Ohnesorge number $\Oh$. In the low-gravity limit, increasing $\Oh$ reduces the take-off velocity according to $U_j/\vgamma \propto (1-\Oh/\Oh_c)$, with $\vgamma=\sqrt{\gamma/(\rho R)}$ for an initial drop radius $R$ and a critical Ohnesorge number $\Oh_c\simeq0.2$ at $\Bo=0$. On the other hand, introducing gravity gives the second arrest mechanism.
Increasing the Bond number $\Bo$ (dimensionless weight), the combined phase boundary for equal drops on an ideal non-wetting substrate is $\Oh/0.2+\Bo/0.9=1$. We also rationalise the observed non-monotonic dimensional velocity, namely, that the small drops are viscously arrested, intermediate drops follow the inertio-capillary scale, and large drops are held down by weight. Coalescence-induced jumping is therefore the surface-energy-driven limit of substrate-mediated capillary recoil, with viscosity and gravity as the two essential controls.
\end{abstract}

\maketitle

\section{Introduction}\label{sec:introduction}

Droplet coalescence on a non-wetting substrate can remove liquid without external forcing \cite{boreyko2009self}. When two neighbouring drops touch, a capillary bridge pulls liquid inward. The substrate blocks the downward part of this motion and redirects part of the impulse upward; if the recoil prevails over dissipation and weight, the coalesced drop jumps.
Jumping renews the substrate, after which condensation of vapour out of a supersaturated environment can lead to the nucleation and growth of subsequent droplets. During dropwise condensation and water harvesting, jumping sustains operation \cite{miljkovic2013jumping,wen2017hierarchical,xie2019dropwise}; the same removal mechanism underpins anti-fogging \cite{mouterde2017antifogging}, self-cleaning \cite{wisdom2013selfcleaning}, and pathogen transport \cite{nath2019sneezing}.
Water-shedding leaves pose the same local problem. Which combination of viscosity and weight allows for droplet take-off?

Before contact, two equal spherical drops have more liquid--gas surface area than a single drop of the same volume. Coalescence therefore releases surface energy, but only the part that survives viscous dissipation, shape oscillations, and gravitational work can become centre-of-mass kinetic energy \cite{boreyko2009self,liu2014numerical,mouterde2017merging,lecointre2019ballistics,wasserfall2017coalescence}. Lecointre \textit{et al.} \cite{lecointre2019ballistics} found that the dimensional jumping velocity first rises and then falls as $R$ varies over several decades. The small-radius branch is viscously dominated whereas the large-radius branch reflects both the decreasing inertio-capillary velocity scale and the increasing weight. Together, these two cut-offs determine the phase space of coalescing droplets.

The corresponding parameter space is governed by
\begin{equation}
    \Oh = \frac{\mu}{\sqrt{\rho\gamma R}} \qquad \text{and} \qquad
    \Bo = \frac{\rho gR^2}{\gamma},
    \label{eq:controls_intro}
\end{equation}
where $R$ is the initial radius of each drop, and $\rho$, $\mu$, $\gamma$, and $g$ are the liquid density, dynamic viscosity, liquid--gas surface tension, and gravitational acceleration. The Ohnesorge number expresses viscous resistance relative to inertio-capillary motion, and the Bond number measures weight relative to capillarity. We use $(\Oh,\Bo)$ to classify both the jump/no-jump outcome and, for jumping cases, the take-off velocity $U_j$.

The same late-recoil balance appears in impact-induced bouncing on non-wetting substrates. In that problem, the Weber number controls the preceding spreading stage, but rebound still fails when capillary recoil cannot overcome viscous losses and weight. The transition follows a capillary--viscous--gravity balance of the form $\Oh/\Oh_c+\Bo/\Bo_c\sim1$ \cite{sanjay2023bouncing}. The comparison with coalescence-induced jumping begins after impact has created the relevant energy reservoir: an impacting drop first spreads and stores capillary energy before recoil, whereas coalescence releases that energy directly as the bridge grows. After that point, capillarity drives radial flow, viscosity drains it, and the substrate redirects the coalesced drop in the upward direction.

This paper quantitatively tests the analogy for two identical drops on an ideal non-wetting substrate. Three-dimensional direct numerical simulations and scaling arguments give the $(\Oh,\Bo)$ phase diagram and the corresponding velocity expression. For negligible gravity, the normalised take-off velocity approaches
\begin{equation}
    \frac{U_j}{\vgamma} \simeq 0.25, \qquad \vgamma=\sqrt{\frac{\gamma}{\rho R}},
    \label{eq:plateau_intro}
\end{equation}
then decreases approximately linearly with $\Oh$ until the viscous cut-off at $\Oh_c(\Bo=0)\simeq0.2$. Gravity sets the second cut-off, $\Bo_c(\Oh\to 0)\simeq0.9$. The transition between jumping and non-jumping takes place at $\Oh/\Oh_c(\Bo=0)+\Bo/\Bo_c(\Oh\to 0)=1$. The two intercepts fix the coefficients. The full phase boundary follows from the same balance that sets the velocity.

The paper is organised as follows. Section~\ref{sec:problem} defines the problem, and Section~\ref{sec:numerics} describes the numerical framework. Section~\ref{sec:phenomenology} discusses the coalescence dynamics and representative jumping/non-jumping scenarios; Section~\ref{sec:velocity} establishes the low-$\Bo$ velocity scaling and validates the dimensional velocity against the measurements of Lecointre \textit{et al.} \cite{lecointre2019ballistics}. Section~\ref{sec:scaling} derives the phase boundary from the capillary--viscous--gravity balance, Section~\ref{sec:energy} analyses the energy budget and the dissipative anomaly in the low-$\Oh$ limit, and Section~\ref{sec:discussion} summarises the results, connects them to impact-induced bouncing, and discusses the limitations of the present work.

\section{Problem formulation}\label{sec:problem}

\begin{figure}[t]
    \centering
    \includegraphics[width=0.8\textwidth]{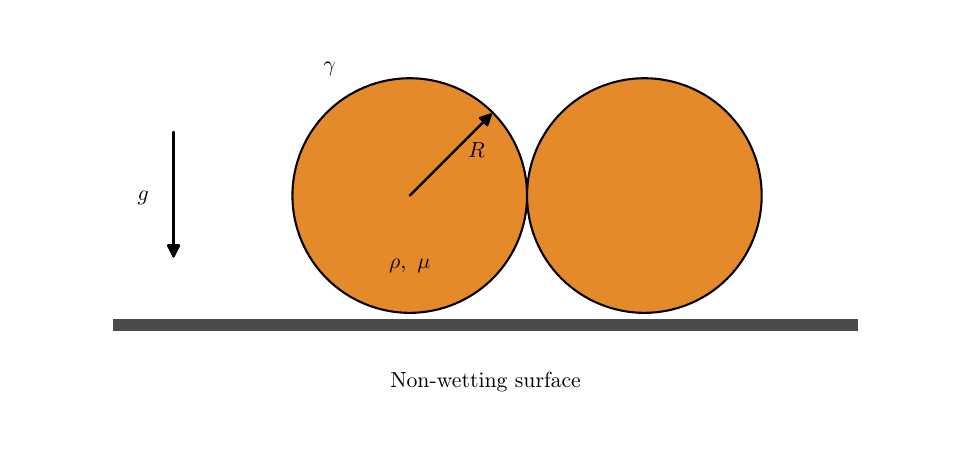}
    \caption{Two identical liquid drops of initial radius $R$, density $\rho$, viscosity $\mu$, and surface tension $\gamma$ coalesce on a non-wetting substrate. Gravity acts downward (acceleration due to gravity is $g$). The problem is governed by $\Oh=\mu/\sqrt{\rho\gamma R}$ and $\Bo=\rho gR^2/\gamma$, and the principal responses are the take-off velocity $U_j$ and the jump/non-jump outcomes.}
    \label{fig:problem_formulation}
\end{figure}

As initial condition, we consider two identical drops of radius $R$ resting on a flat non-wetting substrate, as sketched in Fig.~\ref{fig:problem_formulation}. The drops are Newtonian, incompressible, and surrounded by a passive gas. The liquid has density $\rho$, dynamic viscosity $\mu$, and liquid--gas surface tension $\gamma$. Gravity acts downward with acceleration $g$. The substrate is idealised as non-wetting: it supports the drops while preventing liquid penetration and maintaining a gas layer between the liquid and the solid. This idealisation removes contact-line pinning and isolates the capillary--viscous--gravity mechanism which is the focus of this work.

The inertio-capillary velocity and time scales are
\begin{equation}
    \vgamma=\sqrt{\frac{\gamma}{\rho R}}, \qquad
    \taugamma=\sqrt{\frac{\rho R^3}{\gamma}}=\frac{R}{\vgamma}.
    \label{eq:scales}
\end{equation}
The two control parameters follow from these scales as given in Eq.~\eqref{eq:controls_intro}. In the absence of gravity and viscosity, $\vgamma$ is the only velocity scale. Therefore $U_j/\vgamma$ should be order unity unless the dynamics select an efficiency smaller than one. The simulations and experiments show that they do: in the small-$\Oh$ and small-$\Bo$ limit, the plateau value is approximately $0.25$.

We define a trajectory as jumping if the coalesced drop loses contact with the substrate and, at that instant, its centre of mass is moving upward. The take-off velocity $U_j$ is the vertical centre-of-mass velocity at detachment. Numerically, detachment is identified from two diagnostics: the minimum liquid--substrate separation and the integrated substrate force.

We note that past experiments \cite{boreyko2009self,mouterde2017merging,lecointre2019ballistics} have swept across the parameter space by changing $R$ in a fixed liquid. This would move the system along both axes at once: $\Oh\propto R^{-1/2}$ decreases, whereas $\Bo\propto R^2$ increases. The small-drop limit is viscous; the large-drop limit is gravitational. This opposing dependence is the origin of the non-monotonic dimensional velocity and the basis for the phase diagram derived below.

\section{Numerical framework}\label{sec:numerics}

We simulate the coalescence process using the incompressible Navier--Stokes equations with a geometric volume-of-fluid method \cite{tryggvason2011direct,popinet2009accurate}. In dimensional form, the one-fluid equations read
\begin{align}
    \nabla\cdot \bm{u} &=0, \label{eq:continuity_dim}\\
    \frac{\partial (\rho_f \bm{u})}{\partial t}+\nabla\cdot(\rho_f \bm{u}\bm{u})
    &= -\nabla p + \nabla\cdot\left(2\mu_f\bm{\mathcal{D}}\right)+\bm{f}_{\gamma}-\rho_f g\bm{e}_z,
    \label{eq:momentum_dim}
\end{align}
where $\bm{u}$ is the velocity field, $p$ the pressure, $\bm{\mathcal{D}}=(\nabla\bm{u}+\nabla\bm{u}^{T})/2$ the rate-of-deformation tensor, $\bm{f}_{\gamma}$ the singular capillary force density at the liquid--gas interface, and $\bm{e}_z$ the upward unit vector. The interface is tracked by a volume fraction field $c$, with $c=1$ in the liquid and $c=0$ in the gas, advected by
\begin{equation}
    \frac{\partial c}{\partial t}+\bm{u}\cdot\nabla c=0.
    \label{eq:vof}
\end{equation}
The material properties vary with $c$ through the standard one-fluid approximations, $\rho_f(c)=c\rho+(1-c)\rho_g$ and $\mu_f(c)=c\mu+(1-c)\mu_g$, where $\rho_g$ and $\mu_g$ are the gas density and viscosity.

Using $R$, $\vgamma$, $\taugamma$, and $\gamma/R$ to scale length, velocity, time, and pressure, respectively, gives
\begin{align}
    \tilde{\nabla}\cdot\tilde{\bm{u}} &=0,\label{eq:continuity_nd}\\
    \frac{\partial \tilde{\bm{u}}}{\partial \tilde{t}}+\tilde{\nabla}\cdot(\tilde{\bm{u}}\tilde{\bm{u}})
    &= -\tilde{\nabla}\tilde{p} + \tilde{\nabla}\cdot\left(2\Oh\tilde{\bm{\mathcal{D}}}\right)+\tilde{\bm{f}}_{\gamma}-\Bo\bm{e}_z,
    \label{eq:momentum_nd}
\end{align}
for the liquid phase, with the corresponding density and viscosity ratios retained in the gas through the one-fluid formulation.
The simulations use \textsc{Basilisk C} with geometric VoF interface reconstruction and height-function curvature estimation \cite{popinet2009accurate,popinet2013basilisk,tryggvason2011direct,brackbill1992continuum}. Adaptive mesh refinement concentrates resolution at the interface (tolerance: $10^{-3}$ for $c$ and $10^{-4}$ for curvature $\kappa$), and in regions of large velocities (tolerance: $10^{-2}$). For further details of the code used in this work, see \cite{sanjay2025jumpingdrops}.

The simulations in this work span approximately $10^{-3}\le \Oh\le 0.5$ and $0\le \Bo\le 2$. The initial condition is a pair of equal drops placed in near contact, connected by a tiny liquid bridge that starts the coalescence process. This regularises the mathematical singularity \cite{eggers2025coalescence,sanjay2026singularities} at first contact while preserving the subsequent inertio-capillary dynamics. The simulations are performed in a three-dimensional geometry. The substrate is modelled as impermeable and non-wetting, with a thin gas layer separating the liquid from the substrate.

The additional energy-budget comparison at $\Bo=0$ extends down to $\Oh=10^{-5}$ and includes maximum refinement levels 8, 9, and 10; its resolution limits are discussed in Section~\ref{sec:energy}.

\section{Phenomenology of coalescence-induced jumping}\label{sec:phenomenology}

The motion begins with bridge formation. The large neck curvature produces a Laplace-pressure gradient that draws liquid into the bridge. For freely suspended drops this inward momentum would be redistributed symmetrically. The substrate blocks downward motion, directing part of the flow upward. The coalesced drop flattens, contracts laterally, and recoils into a vertically elongated shape. Whether it takes off depends on the impulse remaining after viscous losses and the work against gravity. We categorize a case as jumping if this impluse is sufficient to detach the lower interface.

\begin{figure}[!htbp]
    \centering
    \includegraphics[width=\textwidth]{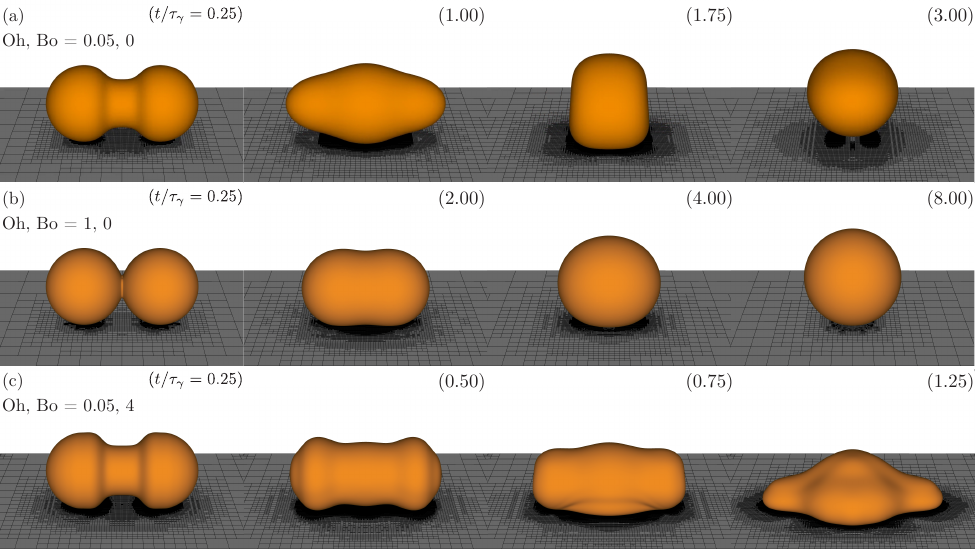}
    \caption{Phenomenology and arrest pathways for coalescence-induced jumping. (a) At low viscosity and negligible gravity, the coalesced drop contracts laterally, recoils vertically, and detaches. (b) At large $\Oh$, viscous stresses damp the recoil before take-off. (c) At large $\Bo$, recoil persists but the drop remains held down by its weight. Times are scaled by $\taugamma=\sqrt{\rho R^3/\gamma}$.}
    \label{fig:phenomenology}
\end{figure}

Fig.~\ref{fig:phenomenology} illustrates this sequence into one jumping trajectory and two arrest pathways. In the low-viscosity, low-gravity case (Fig.~\ref{fig:phenomenology}(a)), the coalesced drop forms a horizontally elongated shape, contracts laterally, recoils upward, and then detaches. The important qualitative observation is that take-off does not yet occur at the instant of first contact between the neck and the substrate. In all three cases (jumping, gravity suppression, viscosity suppression), the two droplets form a bridge and undergo lateral contraction; the outcome is determined later, when the recoil must be converted into centre-of-mass motion. The viscous arrest pathway is shown in Fig.~\ref{fig:phenomenology}(b). Here $\Oh=1$ and $\Bo=0$, so gravity is absent but the internal motion is strongly damped. After coalescence, the coalesced drop becomes spherical owing to surface tension, but the recoil is too slow to produce a finite upward impulse at the substrate. On the other hand, the gravitational arrest pathway is identical to the base case at early times (see $t = 0.25\tau_\gamma$ in Fig.~\ref{fig:phenomenology}(c)). Indeed, at $\Oh=0.05$, the liquid remains mobile, yet $\Bo=4$ makes lifting the coalesced drop too costly: the interface recoils and deforms, but the lower surface stays trapped near the gas layer. Thus both viscosity and weight suppress jumping, but in visibly different ways even though both drive $U_j$ to zero at the transition. Viscous stresses dissipate the excess energy inside the drop during bridge growth and recoil. Gravity instead stores part of the excess energy as gravitational potential energy, so the same recoil can occur without detachment.

\section{Jumping velocity in the low-gravity limit}\label{sec:velocity}

\begin{figure}[t!]
    \centering
    \includegraphics[width=0.85\textwidth]{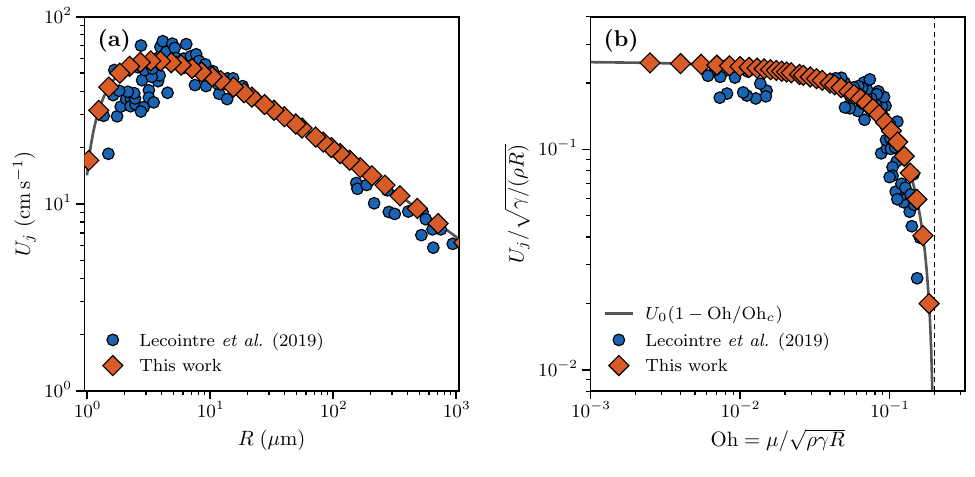}
    \caption{Low-$\Bo$ take-off velocity from the present simulations and measurements of Lecointre \textit{et al.} \cite{lecointre2019ballistics}: (a) $U_j$ against the initial drop radius $R$; (b) $U_j/\vgamma$ against $\Oh$. The solid curve shows Eq.~\eqref{eq:velocity_oh}.}
    \label{fig:velocity_scaling}
\end{figure}

We first discuss the limit in which gravity is negligible ($\Bo \ll 1$). The only relevant velocity scale is then $\vgamma=\sqrt{\gamma/(\rho R)}$. Fig.~\ref{fig:velocity_scaling}(a) compares the dimensional take-off velocity from the simulations with the measurements of Lecointre \textit{et al.} \cite{lecointre2019ballistics}. The data reproduce the non-monotonic behaviour. For small $R$, the velocity increases with increasing radius because $\Oh$ decreases and, consequently, viscous losses weaken. For intermediate radii, the motion is inertio-capillary and $U_j$ follows the $R^{-1/2}$ scaling implicit in $\vgamma$. We note that, in this figure, $R \lesssim \sqrt{\gamma/(\rho g)} = 2.7\,\text{mm}$; for still larger radii, gravity becomes relevant through $\Bo\propto R^2$ and ultimately prevents jumping (see Section~\ref{sec:scaling}).

The collapse becomes clearer after non-dimensionalising the velocity by $\vgamma$ and plotting it against $\Oh$, as in Fig.~\ref{fig:velocity_scaling}(b). In the range $\Bo\ll1$, the data follow
\begin{equation}
    \frac{U_j}{\vgamma} \simeq U_0\left(1-\frac{\Oh}{\Oh_c(\Bo=0)}\right), \qquad
    U_0\simeq0.25, \qquad \Oh_c(\Bo=0)\simeq0.2.
    \label{eq:velocity_oh}
\end{equation}
Capillarity sets the velocity, but the prefactor is not unity (see Section~\ref{sec:energy}). Only a fraction of the released surface energy and radial momentum becomes centre-of-mass translation. The rest remains in shape oscillations or is internally dissipated.

Viscous stresses dissipate energy during bridge growth and recoil. For an inertio-capillary flow with velocity gradients of order $\vgamma/R$ over a volume of order $R^3$, the drop-scale contribution to the dissipated power is
\begin{equation}
    \mathcal{P}_\eta \sim \mu\left(\frac{\vgamma}{R}\right)^2 R^3.
    \label{eq:viscous_power}
\end{equation}
We express this loss as an effective resistance to the internal flow, $F_\eta\vgamma\sim\mathcal{P}_\eta$, giving
\begin{equation}
    F_\eta \sim \frac{\mathcal{P}_\eta}{\vgamma} \sim \mu \frac{\vgamma}{R^2}R^3 \sim \mu\vgamma R.
    \label{eq:viscous_force}
\end{equation}
The same scaling follows from the viscous stress acting over a drop-scale area, $F_\eta\sim\mu(\vgamma/R)R^2$. Compared with the capillary drive $F_\gamma\sim\gamma R$, this gives $F_\eta/F_\gamma\sim\Oh$. To relate this resistance to the take-off velocity, we use the flow-redirection model in Section~\ref{sec:scaling}. Taking the recoil duration to scale with $\taugamma$ and the redirection efficiency to be independent of $\Oh$ at leading order gives a linear viscous correction to the upward impulse, and hence to $U_j$. These assumptions motivate Eq.~\eqref{eq:velocity_oh}; its extension to the viscous cut-off is calibrated by the data. The drop-scale estimate does not include the dissipation associated with gradients that sharpen as $\Oh\to0$, discussed in Section~\ref{sec:energy}.

At the cut-off measured (through the DNS), $\Oh_c(\Bo=0)\simeq0.2$, viscous stress dissipates the coalescence-induced motion before take-off. This value is lower than the order-one criterion one would obtain from a crude dimensional balance. The reason is that the conversion from capillary energy to centre-of-mass kinetic energy is already inefficient in the low-viscosity limit. The same inefficiency appears in the plateau value $U_0\simeq0.25$.

\section{Scaling argument and phase boundary}\label{sec:scaling}

\begin{figure}[t]
    \centering
    \includegraphics[width=0.8\textwidth]{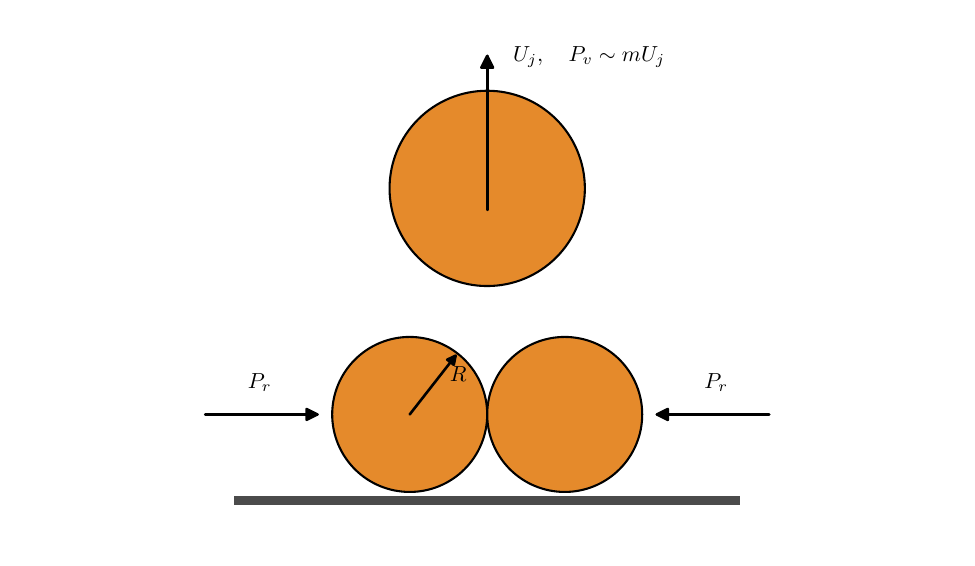}
    \caption{Substrate-mediated redirection of the capillary flow. $P_r$ is the inward momentum scale and $P_v\sim mU_j$ the vertical momentum, where $m$ is the coalesced-drop mass.}
    \label{fig:scaling_schematic}
\end{figure}

Bridge growth creates an inward capillary flow, which viscous dissipation damps. Here $P_r$ denotes the characteristic magnitude of the inward momentum carried by this flow; the opposing horizontal contributions cancel in the total momentum of the two-drop system. The substrate blocks downward motion and supplies an upward reaction through the gas layer. This external force changes the total vertical momentum $P_v$, while viscous dissipation reduces the internal motion that drives the reaction. As sketched in Fig.~\ref{fig:scaling_schematic}, we model this redirection by
\begin{equation}
    P_v \sim P_r.
    \label{eq:momentum_redirection}
\end{equation}
This is the usual closure in coalescence-induced jumping models and in bouncing-to-non-bouncing criteria for impacting drops \cite{liu2014numerical,mouterde2017merging,lecointre2019ballistics,sanjay2023bouncing}. Assuming that the two momentum scales vary over the same recoil time, with a constant redirection factor, Eq.~\eqref{eq:momentum_redirection} gives the corresponding estimate for the magnitudes of their characteristic rates of change,
\begin{equation}
    |\dot{P}_v| \sim |\dot{P}_r|.
    \label{eq:force_redirection}
\end{equation}
We use this closure during recoil to estimate the jump/no-jump threshold. When the effective capillary drive remaining after viscous losses can no longer overcome the weight of the coalesced drop,
\begin{equation}
    F_\gamma - F_\eta \sim F_g.
    \label{eq:threshold_force_balance}
\end{equation}

\begin{figure}[t]
    \centering
    \includegraphics{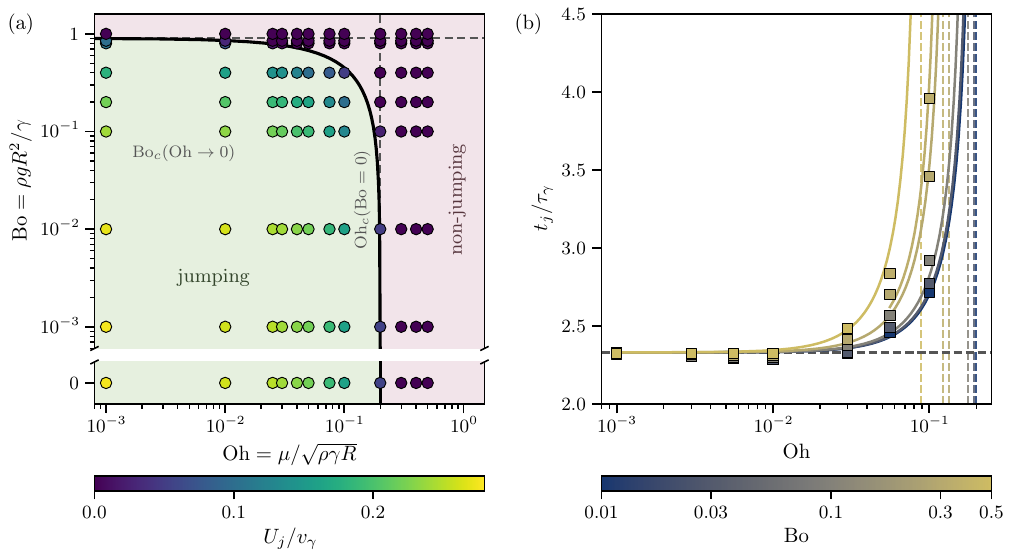}
    \caption{(a) Jumping/non-jumping phase map, coloured by $U_j/\vgamma$. The solid black curve is the fitted boundary, Eq.~\eqref{eq:phase_boundary}; dashed lines mark the limiting cut-offs $\Oh_c(\Bo=0)$ and $\Bo_c(\Oh\to 0)$. The lower strip shows $\Bo=0$. (b) Take-off time: squares show DNS, coloured by $\Bo$; solid curves show Eq.~\eqref{eq:takeoff_time}. Horizontal and vertical dashed lines mark the low-$\Oh$ time and $\Oh_c(\Bo)$, respectively.}
    \label{fig:phase_map}
\end{figure}

Equation~\eqref{eq:threshold_force_balance} is an effective recoil balance: $F_\gamma$ and $F_\eta$ describe the internal drive and resistance coupled to vertical motion by the substrate. Estimating these scales and the weight over the drop scale gives $F_\gamma\sim\gamma R$, $F_\eta\sim\mu\vgamma R$, and $F_g\sim\rho gR^3$. The dimensional threshold balance therefore becomes
\begin{equation}
    \gamma R - C_\eta \mu\vgamma R \sim C_g \rho gR^3,
    \label{eq:force_balance_dim}
\end{equation}
where $C_\eta$ and $C_g$ contain the geometric and redirection factors, normalized by the recoil drive that survives in the low-$\Oh$ limit. The finite dissipation discussed in Section~\ref{sec:energy} reduces this reference drive; the term proportional to $C_\eta\Oh$ describes the additional drop-scale viscous resistance. Dividing Eq.~\eqref{eq:force_balance_dim} by $\gamma R$ gives
\begin{equation}
    C_\eta \Oh + C_g \Bo \sim 1.
    \label{eq:force_balance_nd}
\end{equation}
We use Eq.~\eqref{eq:force_balance_nd} to constrain the boundary to a straight line in the $(\Oh,\Bo)$ plane and fit its position by tracing the transition between the jumping and non-jumping DNS outcomes in Fig.~\ref{fig:phase_map}(a). The two fitted axis intercepts are $\Oh_c(\Bo=0)\simeq0.2$ and $\Bo_c(\Oh\to 0)\simeq0.9$. Consequently, $C_\eta\Oh_c(\Bo=0)=1$ and $C_g\Bo_c(\Oh\to 0)=1$, so that $C_\eta=1/\Oh_c(\Bo=0)$ and $C_g=1/\Bo_c(\Oh\to 0)$. Substitution into Eq.~\eqref{eq:force_balance_nd} gives
\begin{equation}
    \frac{\Oh}{\Oh_c(\Bo=0)}+\frac{\Bo}{\Bo_c(\Oh\to 0)}=1,
    \qquad \Oh_c(\Bo=0)\simeq0.2, \qquad \Bo_c(\Oh\to 0)\simeq0.9.
    \label{eq:phase_boundary}
\end{equation}
This relation is the proposed jump/no-jump boundary.

The phase boundary specifies where $U_j$ vanishes but does not by itself determine the velocity away from it. Motivated by the approximately linear low-$\Bo$ dependence in Eq.~\eqref{eq:velocity_oh}, we use a linear function of the signed distance from the two-parameter boundary as a first-order near-threshold extension,
\begin{equation}
    \frac{U_j}{\vgamma}\simeq U_0\left(1-\frac{\Oh}{\Oh_c(\Bo=0)}-\frac{\Bo}{\Bo_c(\Oh\to 0)}\right),
    \label{eq:velocity_two_parameter}
\end{equation}
which reduces to Eq.~\eqref{eq:velocity_oh} when $\Bo\ll1$. Equation~\eqref{eq:velocity_two_parameter} is valid near the transition and does not account for shape-oscillations away from this transition. We can recover Eq.~\eqref{eq:phase_boundary} by setting $U_j=0$.

Fig.~\ref{fig:phase_map}(a) compares this boundary with the numerical phase map. Drops jump only in the low-$\Oh$, low-$\Bo$ corner. At fixed $\Bo$, increasing $\Oh$ damps the recoil; at fixed $\Oh$, increasing $\Bo$ can prevent lift-off even when recoil persists.

The two intercepts quantify the efficiency of converting the capillary release into upward centre-of-mass motion. If that conversion were perfect, the gravitational cut-off would occur at the crude balance $F_g\sim F_\gamma$, i.e. $\Bo_c(\Oh\to 0)\sim1$. The simulations give $\Bo_c(\Oh\to 0)\simeq0.9$ because, even as $\Oh\to0$, part of the released surface energy remains in internal motion and shape deformation, and part is lost through viscous dissipation, which may remain finite as $\Oh\to0^+$ (see Section~\ref{sec:energy}). The same accounting explains $\Oh_c(\Bo=0)\simeq0.2$. The estimate $F_\eta/F_\gamma\sim\Oh$ compares viscous resistance with the total capillary drive; arrest requires viscosity to drain only the smaller impulse that the substrate would otherwise redirect into upward motion. The viscous threshold is therefore well below unity.

The take-off time $t_j$ is defined by the detachment criterion in Section~\ref{sec:problem}. Figure~\ref{fig:phase_map}(b) shows nearly constant times at low $\Oh$ and delayed take-off towards the jumping boundary. To test the jumping--bouncing analogy, we apply the spring--mass--damper contact-time relation for viscous bouncing drops \cite{jha2020viscous,sanjay2023bouncing} to the take-off time:
\begin{equation}
    \frac{t_j}{\taugamma}\simeq 2.33\left[1-\left(\frac{\Oh}{\Oh_c(\Bo)}\right)^2\right]^{-1/2},
    \qquad \Oh_c(\Bo)=\Oh_c(\Bo=0)\left(1-\frac{\Bo}{\Bo_c(\Oh\to 0)}\right).
    \label{eq:takeoff_time}
\end{equation}
The low-$\Oh$ asymptote is fitted to the take-off times, while $\Oh_c(\Bo)$ follows from the transition boundary in Eq.~\eqref{eq:phase_boundary}. For $\Bo\le0.5$ and $\Oh\le0.1$, the simulated times lie within $6\%$ of the prediction.

\section{Energy budget and dissipative anomaly}\label{sec:energy}

The low-$\Oh$ plateau in both simulations and experiments \cite{lecointre2019ballistics}, $U_j/\vgamma\simeq0.25$, shows that only a small fraction of the released surface energy drives centre-of-mass translation. We note that a low take-off velocity alone does not distinguish viscous dissipation from energy retained in internal motion and shape deformation. We therefore examine these contributions separately. A dissipative anomaly would arise if, as $\Oh\to0^+$, velocity gradients sharpened sufficiently to keep the dissipation rate $\nu\langle(\partial_i u_j)^2\rangle_V$ finite and nonzero, where $\nu = \mu/\rho$ is the kinematic viscosity.

To analyse this limit, we decompose the energy into surface, kinetic, and dissipated contributions (for $\Bo = 0$),
\begin{align}
    E_s(t) &= \int_A \gamma\,\dd A {}-4\pi\gamma\left(2^{1/3}R\right)^2, \label{eq:surface_energy}\\
    E_k(t) &= \int_V \frac{1}{2}\rho_f(c) |\bm{u}|^2\,\dd V, \label{eq:kinetic_energy}\\
    E_d(t) &= \int_0^t\int_V 2\mu_f(c)\bm{\mathcal{D}}:\bm{\mathcal{D}}\,\dd V\,\dd t'. \label{eq:dissipation_energy}
\end{align}
The volume integrals include both liquid and gas, with the phase-dependent density $\rho_f(c)$ and viscosity $\mu_f(c)$ defined in Section~\ref{sec:numerics}. Thus $E_k=E_k^{\ell}+E_k^g$ and $E_d=E_d^{\ell}+E_d^g$, where the superscripts denote the liquid and gas contributions.
We have taken the surface energy $4\pi\gamma(2^{1/3}R)^2$ of the final spherical drop as the reference level. Thus $E_s$ measures the excess surface energy available for release, whose initial value is denoted by $E_0$.

\begin{figure}[t]
    \centering
    \includegraphics[width=\textwidth]{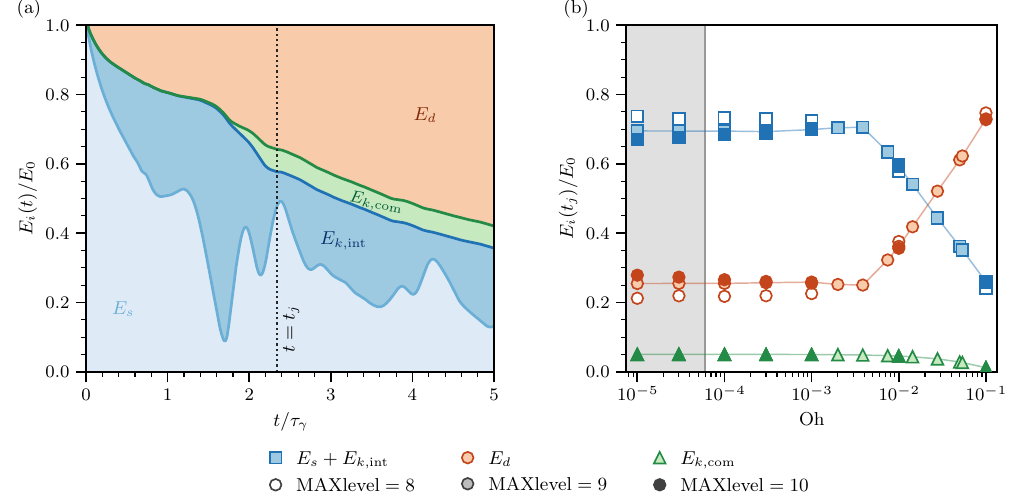}
    \caption{Energy fractions at $\Bo=0$, normalized by the initial excess surface energy $E_0$. (a) Stacked budget at $\Oh=0.01$; the dotted vertical line marks take-off. (b) Take-off partition for $\Oh\geq10^{-5}$; lines connect MAXlevel~9 data. The grey region, $\Oh<\Oh^*\simeq6\times10^{-5}$, marks the range where physical and numerical dissipation cannot be distinguished reliably.}
    \label{fig:energy_budget}
\end{figure}

Figure~\ref{fig:energy_budget}(a) shows the energy budget at $\Oh=0.01$, normalised by the initial excess surface energy $E_0$. Panel (b) separates the centre-of-mass kinetic energy $E_{k,\mathrm{com}}=\tfrac{1}{2}mU_j^2$ from the internal kinetic energy $E_{k,\mathrm{int}}=E_k-E_{k,\mathrm{com}}$ at take-off in the completed $\Bo=0$ simulations. Here $m$ is the liquid mass; $E_{k,\mathrm{int}}$ includes internal liquid motion and the kinetic energy of the surrounding gas.
As the bridge grows, surface energy decreases and kinetic energy rises, subsequently feeding shape oscillations; viscous dissipation accumulates throughout. Dissipation at finite $\Oh$ is not by itself anomalous. The analogous bouncing problem motivates testing whether a finite dissipated-energy fraction persists as $\Oh\to0^+$ \cite{sanjay2023bouncing,sanjay2025unifying}.
Figure~\ref{fig:phase_map}(b) shows that $t_j$ is nearly independent of $\Oh$ and $\Bo$ for $\Bo\le0.1$ and $10^{-3}\le\Oh\lesssim0.03$; over this range, Fig.~\ref{fig:energy_budget}(b) compares cases over nearly equal recoil times. The plateau $E_d(t_j)/E_0\simeq0.25$ between $\Oh=10^{-3}$ and $4\times10^{-3}$ thus implies a nearly constant time-averaged dissipation rate $E_d(t_j)/t_j$. For $\Oh$-independent velocity gradients, a regular viscous contribution would instead decrease in proportion to $\Oh$. At larger $\Oh$, $E_d(t_j)/E_0$ reaches about one half near $\Oh=0.03$ while $t_j$ remains close to its low-$\Oh$ value, indicating a larger dissipation rate rather than a longer recoil.

Refinement does not remove the finite low-$\Oh$ loss in Fig.~\ref{fig:energy_budget}(b). At $\Oh=10^{-3}$, MAXlevel~9 and 10 give $E_d(t_j)/E_0\simeq0.259$ and $0.256$, respectively, but the lower-$\Oh$ fractions remain more sensitive to resolution. Convergence of volume-integrated momentum does not ensure convergence of dissipation, which depends on squared velocity gradients in thin viscous layers. Stevens \textit{et al.}~\cite{stevens2010radial} illustrate this sensitivity to under-resolution in Rayleigh--B\'enard convection. Using the boundary-layer resolution criterion of Kayal \textit{et al.}~\cite{kayal2025focusing}, we estimate a lower limit $\Oh^*\simeq6\times10^{-5}$, below which physical and numerical dissipation cannot be distinguished reliably (see the shaded region in Fig.~\ref{fig:energy_budget}(b)). The finite viscous plateau on the finer grids supports the proposed dissipative anomaly, but does not establish whether the dissipated fraction approaches a finite nonzero limit as $\Oh\to0^+$.

We also note that these sinks are not distributed uniformly through the drop. At first contact, the small neck length scale produces large velocity gradients. Capillary waves travel over the free surface, leaving thin viscous layers in the liquid and gas, while residual shape oscillations exchange surface and internal kinetic energy before take-off. In each region, vanishing values of $\Oh$ in the dimensionless viscous dissipation rates $\tilde{\varepsilon}_\mu = 2\Oh\bm{\tilde{\mathcal{D}}}:\bm{\tilde{\mathcal{D}}}$ can be compensated by gradients that sharpen as $\mu$ decreases. This is the same singular mechanism class seen in retracting sheets, free-surface wave focusing, and low-viscosity drop rebound \cite{sanjay2022taylor,kayal2025focusing,sanjay2025unifying}. Finite viscous losses, together with energy retained in internal motion and deformation, account for the low-$\Oh$ velocity plateau $U_0\simeq0.25$.

\section{Conclusion and outlook}\label{sec:discussion}

This work elucidates how viscous dissipation and weight suppress the coalescence-induced jumping of two identical drops \cite{boreyko2009self}. As the two drops coalesce, the liquid bridge releases surface energy and drives inward capillary flow. Viscosity drains part of that energy while the substrate redirects the remaining radial momentum upward. Consequently, gravity sets the work required to lift the coalesced drop.
Drops smaller than their visco-capillary length, i.e., $\Oh \gtrsim \mathcal{O}(1)$, do not jump because viscous stresses dissipate the coalescence-induced motion, whereas drops larger than their gravito-capillary length, i.e., $\Bo \gtrsim \mathcal{O}(1)$, do not jump because their weight exhausts the available capillary work.
These two limits are the coalescence analogue of the bouncing/non-bouncing transition for impacting drops \cite{sanjay2023bouncing}.

For two equal drops on an ideal non-wetting substrate, the low-gravity velocity follows
\begin{equation}
    \frac{U_j}{\sqrt{\gamma/(\rho R)}}\simeq0.25\left(1-\frac{\Oh}{0.2}\right),
\end{equation}
so the low-$\Oh$ limit is a plateau of $0.25$, not an inviscid conversion efficiency near unity. Adding gravity gives the jump/no-jump transition
\begin{equation}
    \frac{\Oh}{0.2}+\frac{\Bo}{0.9}=1.
\end{equation}
Here, we stress that while the two intercepts come from the two physical asymptotes, they are system dependent. They are valid for equal-sized drops on an ideal non-wetting substrate, where small drops fail because viscous dissipation removes too much of the redirected impulse and large drops fail because lifting the coalesced drop consumes the available capillary work.

The dimensional velocity $U_j(R)$ is non-monotonic \cite{lecointre2019ballistics}. Increasing $R$ decreases $\Oh$ and thus weakens viscous losses, initially increasing $U_j$.
At intermediate radii, $U_j$ follows the inertio-capillary scale $\sqrt{\gamma/(\rho R)}$ and decreases with $R$.
For larger $R$, $\Bo$ approaches its critical value and gravity suppresses take-off.
Coalescence-induced jumping is therefore the zero-impact limit of substrate-mediated capillary recoil; viscous dissipation and gravity set the jump/no-jump boundary.
That comparison is only at the level of the late recoil. In impact-induced bouncing, the Weber number $\We$ records the preceding spreading history: it sets the deformation, film thickness, and partitioning of the initial kinetic energy before recoil begins \cite{sanjay2025unifying}. Once the drop recoils, the same capillary, viscous, and gravitational force scales enter as in jumping, but with problem-dependent prefactors and more scatter in the phase boundary \cite{sanjay2023bouncing}. Coalescence-induced jumping removes this impact history and starts directly from a surface-energy reservoir, leaving $\Oh$ and $\Bo$ as the two essential controls.

Finally, we emphasise that the phase boundary described in this work also applies to an ideal non-wetting substrate.
Finite roughness and textured geometry, contact line pinning, and possible contact line hysteresis can further remove the available energy and need to be systematically studied.
In the context of this work, Eq.~\eqref{eq:phase_boundary} should be read as the outer bound of the jumping/non-jumping transition on ideal non-wetting substrates.
Roughness and contact-line physics can then be interpreted as perturbations to $\Oh_c(\Bo=0)$ and $\Bo_c(\Oh\to 0)$, or, if strong enough, as additional failure mechanisms.
By increasing $R$, two opposing effects compete, namely $\Oh$ is lowered but $\Bo$ is raised, so that the trajectory in the $(\Oh,\Bo)$ plane does not isolate either limit.
An experimental low-$\Bo$ viscosity series would test Eq.~\eqref{eq:velocity_oh}; an experimental low-$\Oh$ gravity series using radius, effective gravity, or density contrast would test $\Bo_c(\Oh\to 0)\simeq0.9$.

\section*{Acknowledgements and Funding}
\noindent\textbf{Acknowledgements.} This paper is dedicated to the memory of Andreas (Andy) Acrivos, who was a great man and a giant in fluid dynamics, and who taught us in many ways. We also thank Andrea Prosperetti for illuminating discussions.

\noindent\textbf{Funding.} We acknowledge funding from the NWO-Canon grant FIP-II. This work was carried out on the national e-infrastructure of SURFsara, a subsidiary of SURF cooperation, the collaborative ICT organization for Dutch education and research. This work was sponsored by NWO - Domain Science for the use of supercomputer facilities. V.S. acknowledges start-up funding from Durham University.

\bibliographystyle{apsrev4-2-title}
\bibliography{references}

\end{document}